# Power Control Protocols in VANET

**Ghassan Samara**
*Department of Computer Science*
*Faculty of Science and Information Technology*
*Zarqa University, Zarqa, Jordan*
E-mail: Gsamara@zu.edu.jo

**Amer O Abu Salem**
*Department of Computer Science*
*Faculty of Science and Information Technology*
*Zarqa University, Zarqa, Jordan*
E-mail: amer407@yahoo.com,

**Tareq Alhmiedat**
*Department of Computer Science,*
*Faculty of Science and Information Technology*
*Zarqa University, Zarqa, Jordan*
E-mail: t.alhmiedat@gmail.com

**Abstract**

Vehicular Ad hoc Networks (VANET) is one of the most challenging research area in the field of the Mobile Ad hoc Network (MANET), Power control is a critical issue in VANETwhere is should be managed carefully to help the channel to have high performance.
In this paper a comparative study in the published protocols in the field of safety message dynamic power control will be presented and evaluated

## Introduction

Vehicular Ad hoc Networks (VANET) is part of Mobile Ad Hoc Networks (MANET)Samara et al. (2011) and Samara et al. (2012), see figure 1. This means that every node can move freely within the network coverage and stay connected without wires, each node can communicate with other nodes in single hop or multi hop, and any node could be Vehicle, Road Side Unit (RSU). The main difference between VANET and MANET is that VANET consists of high mobile nodes and usually having dense situations.

**Figure 1:** One of VANET applications.

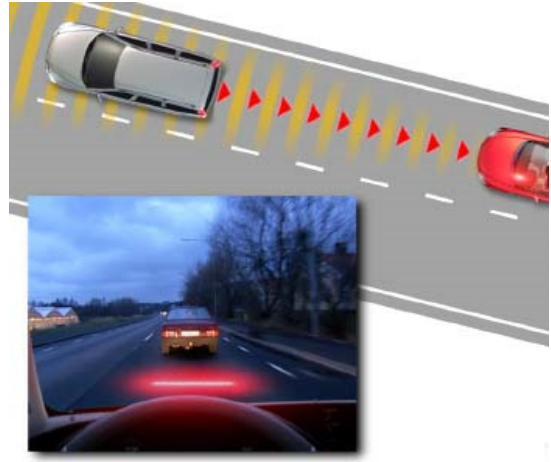

Wirelesses access in vehicular environment (WAVE) is a multi-channel approach, designed by the Federal Communications Commission (FCC), reserved for one control channel from 5.855 to 5865 GHz, for high availability, low latency vehicle safety communications (Commission, 2008). Furthermore, WAVE represents the first VANET standard published in 2006. An enhancement was required on IEEE 802.11 standard to support applications from the Intelligent Transportation Systems (ITS), a branch of the U.S. Department of Transportation. The result showed the 802.11p standard, which was approved on July 2010 (Grouper, 2011). The 802.11p standard is meant for VANET communication and uses dedicated short range communications (DSRC) spectrum; it is divided into eight 10 MHz channels with only one control channel for safety application communication. VANET safety applications depend on the exchange of safety information among vehicles (C2C communication) or between vehicle to infrastructure (C2I Communication) using the control channel (see Figure 2). VANET safety communication is implemented in two ways, namely, periodic safety message (hereby called beacon) and event-driven message (hereby called emergency message), both sharing only one control channel. The beacon messages are messages containing status information about the sender vehicle, such as position, speed, heading, and others. Beacons provide fresh information about the sender vehicle to the surrounding vehicles in the network, updating them of the status of the current network and predicting the movement of vehicles. Beacons are sent aggressively to neighboring vehicles at 10 messages each second. In turn, this causes an increase in channel collision that the control channel cannot tolerate, especially when dense traffic occurs in small geographic areas.

**Figure 2:** VANET Structure

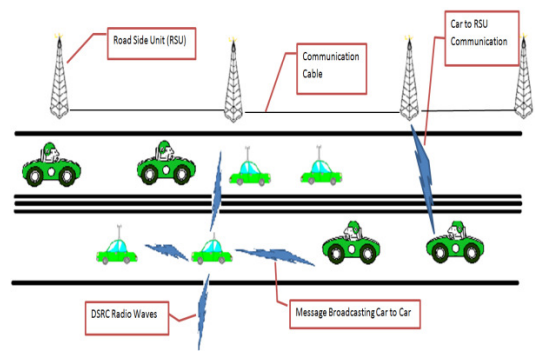

The VANET structure controlling beacon messages could be executed by transmission power control or message repetition control. Sending the message on high full power may cause the message to reach longer distances, thereby increasing the channel load, whereas sending in low power enables the message to reach only very short distances.

## Published Power Control Protocols

Power control in ad hoc networks has been an active topic for many years in the field of topology control. However, vehicular networks' main design goal as a safety system makes all these analyses or proposed algorithms insufficient in satisfying VANET requirements. Most of these studies addressed uni-cast environments and have been intended to improve energy consumption. In the literature, some studies have proposed the best path to the destination that minimize energy consumption and/or maximizes the overall throughput, including those of Kawadia and Kumar (2005), Chen et al. (2003b), and Kubisch et al. (2003).

Chen et al. (2003a) have proposed an "energy aware" adaptive algorithm, which uses only local information to adjust power. Park and Sivakumar (2002), Park and Sivakumar (2003), and Liu and Li (2002) all agree that the minimum transmission power does not always maximize throughput. Although many studies in this field can be found, VANET energy efficiency is not an issue where nodes have a nearly unlimited power supply for communication. Rawat et al. (2011) proposed dynamic adjustment of transmission power based on estimates of local vehicle density. However, traffic density does not indicate channel load; thus, if the channel load is high and the traffic density is low, the sender chooses high power for sending the message, further increasing channel load and causing message reception failure.

Mittag (2009) presented a comparison between single-hop transmission at high transmission power and multi-hop transmission at low transmission power to determine whether or not efficient multi-hop beaconing can reduce channel load. The author found that single hop is best for beaconing and multi hop is best for full coverage. Sending in high power enables beacons to reach long distances in single-hop and may increase channel load. Broadcasting at full power, by comparison, produces a broadcast storm problem (Ni et al., 1999) and raises channel load.

Meanwhile, Guan (2007) developed a power control algorithm to determine optimum transmission power for beacon message transmission by adding a power tuning feedback beacon during each beacon message exchange. On each message exchange, the sender calculates the distance to the receiver and sets a predicted transmission power. On the receiver side, the distance is computed to determine if the transmission power achieved a greater distance or not. However, the delay resulting from these message exchanges makes the information gathered outdated as network status is variable.

Li et al. (2004b) proposed an analytical model to find a transmission power, which maximizes single-hop broadcast coverage. Li et al. (2004c) also proposed an adaptive algorithm that adjusts to a given fixed transmission power. Although both studies focused on a pure broadcast environment, their assumptions made their approach infeasible for vehicular networks, because their nodes were static and had the same priority, i.e., there was no difference between the transmission power of beacon and emergency messages.

Chigan (2007) proposed a Delay-Bounded Dynamic Interactive Power Control (DB-DIPC), in which the transmission powers of VANET nodes are verified by neighboring vehicles at run-time. The idea is to send beacons to neighbor vehicles at very low power, and if the sender receives an acknowledgment, then that specific power is sufficient for close neighbors. This mechanism sends beacons to very close vehicles and limits the information gain for vehicles in the network. It also produces a very long delay as the sender needs to send the message many times to its neighbors and wait for a reply to decide the suitable transmission power.

Torrent-Moreno (2007) proposed the Fair Power Adjustment for Vehicular environments (DFPAV), which tries to adjust the channel load in a VANET environment by maximizing the minimum transmission range for all nodes using a synchronized approach. This is done by analyzing the piggybacked beacon information received from neighbors.

The FPAV protocol is widely recognized for controlling channel load in a fair manner. In this scheme, every node uses a localized algorithm based on a "water filling" approach as proposed by Gallager (1987) and starts transmitting the beacon message with the minimum transmission power. All the nodes increase their transmit power simultaneously to the same maximum power, while the constraint on the beaconing network load MBL is not violated.

According to the analysis of the DFPAV protocol conducted by Mittag (2008), the overhead for the existing DFPAV approach can be reduced, but there is still room for improvement.

Artimy et al. (2005) based transmission range on traffic density estimation, in which an algorithm sets vehicle transmission range dynamically according to local traffic conditions. Artimy, M. M., Rrobertson, W. & Philips, W. J. (2005), Assignment of dynamic transmission range based on estimation of vehicle density, 2nd ACM international workshop on Vehicular ad hoc networks, 40-48, ACM.

This protocol analyzes traffic conditions and not the channel status; hence, the channel may sometimes suffer from collisions when traffic is not dense.

Khorakhun et al. (2008) proposed power control assignment based on network channel busy time as wireless channel quality. When the channel busy time is higher or lower than a desired threshold, specific actions are conducted. However, since threshold selection is arbitrary, outcomes are not always optimal. Table 2.3 compares the previously mentioned protocols in the field of safety message dynamic power control.

Samara and Alsalihy (2010) proposed a dynamic mechanism to control the transmission power for beacon messages depending on the channel status, no experimental results were presented.

## Conclusion

This paper has presented a comparative study in the published protocols in the field of safety message dynamic power control and the main observations are:
1. Energy consumption is not an issue in VANET, as vehicles have rich resources of power.
2. Beacon messages create a growing collision in the control channel.
3. Beacon should be broadcasted in a single-hop to avoid further channel collision.
4. Piggyback the power information used in the transmission into the beacon helps to analyze the network.
5. The DFPAV controls the channel collision by using a dynamic transmission power adjustment depending on the fairness concept.

**Appendix**

Table 1 compares the previously mentioned protocols in the field of safety message dynamic power control.

**Table 1:** Comparison table for protocols in the field of Performance of emergency message system.

| \multicolumn{4}{c}{Safety Message Dynamic Power Control} | | | |
|---|---|---|---|
| **L.R.** | **Methodology** | **Results** | **Weakness** |
| (Li et al., 2004b), (Li et al., 2004c) | Analytical model able to find a transmission power that maximizes single-hop broadcast coverage. | Maximize the broadcast coverage. | a) all nodes are static b) all nodes have the same priority (i.e., no difference between the transmission power for beacon and emergency message). |
| (Mittag, 2009) | Made a comparison between single-hop transmissions at high transmit power and multi-hop broadcasting and rebroadcasting at lower transmission power. | founds that single hop must be used for beaconing | may increase the channel Collision |
| (Chigan, 2007) | Delay-bounded dynamic interactive power control (DB-DIPC), in where vehicle sends message in low power to its neighbors and waits for the acknowledgment to test if this power is enough. | Decide the transmission power for the close neighbors | Very long delay, depends only on very close neighbors. |
| (Guan et al., 2007) | A power control algorithm to determine the transmission power by adding a power tuning feedback beacon during each safety message exchange. | More data traffic loads on the channel, the greater information about the channel. | Adding all the power data into every beacon will increase the channel payload and channel collision |
| (Torrent-Moreno, 2005) | FPAV, a centralized power control algorithm by maximizing the minimum transmission range for all nodes in a synchronized approach, by analyzing the piggybacked beacon information received from neighbors. | Adjust the load on the channel. | No technique for analyzing the channel status, all the nodes must send by the same power |
| (Torrent-Moreno et al., 2006) | Distributed Fair Power Adjustment for Vehicular environments (D-FPAV), all vehicles starts to transmit the beacon messages from the minimum transmission value and all the nodes increase their transmit power simultaneously with the same number of maximum power while the constraint on the beaconing network load MBL is not violated. | Channel collision is reduced. | (Mittag, 2008) made analyses on D-FAPV protocol, the result showed that DFAPV needs improvement in saturated traffic condition. The sender vehicle doesn't depend on the channel status for selecting the suitable power. |
| (Artimy et al., 2005) | Transmission Power selection depending on traffic density estimation. | Considerable increase in message range. | Depends on traffic conditions not the channel status, sometimes the channel suffers from collision while the traffic is not dense. |
| (Khorakhun et al., 2008) | a power value based on a network channel busy time, When the channel busy time is higher or below than a desired threshold. | The protocol outcomes are not always optimal. | Threshold selection is arbitrary. |